\documentclass[aps,twocolumn,showpacs,amsmath,amssymb,floatfix,eqsecnum]{revtex4}
\usepackage{graphicx}
\usepackage{dcolumn}
\usepackage{bm}

\begin{document}

\title{Universality of the thermodynamic Casimir effect}
\author{Daniel Dantchev$^{1,2}$\thanks{e-mail:
danield@mf.mpg.de}, Michael Krech$^{2,3}$\thanks{e-mail:
mkrech@mf.mpg.de}, and S. Dietrich$^{2,3}$\thanks{e-mail:
dietrich@mf.mpg.de}} \affiliation{ $^1$Institute of Mechanics -
BAS, Acad. G. Bonchev St. bl. 4,
1113 Sofia, Bulgaria,\\
$^2$Max-Planck-Institut f\"{u}r Metallforschung,
Heisenbergstrasse 1, D-70569 Stuttgart, Germany,\\
$^3$Institut f\"{u}r Theoretische und Angewandte Physik,
Universit\"{a}t Stuttgart, Pfaffenwaldring 57, D-70569 Stuttgart,
Germany}

\begin{abstract}
Recently a nonuniversal character of the leading spatial behavior
of the thermodynamic Casimir force has been reported [X. S. Chen
and V. Dohm, Phys. Rev. E {\bf 66}, 016102 (2002)]. We reconsider
the arguments leading to this observation and show that there is
no such leading nonuniversal term in systems with short-ranged
interactions if one treats properly the effects generated by a
sharp momentum cutoff in the Fourier transform of the interaction
potential. We also conclude that lattice and continuum models then
produce results in mutual agreement independent of the cutoff
scheme, contrary to the aforementioned report. All results are
consistent with the {\em universal} character of the Casimir force
in systems with short-ranged interactions. The effects due to
dispersion forces are discussed for systems with periodic or
realistic boundary conditions. In contrast to systems with
short-ranged interactions, for $L/\xi \gg 1$ one observes leading
finite-size contributions governed by power laws in $L$ due to the
subleading long-ranged character of the interaction, where $L$ is
the finite system size and $\xi$ is the correlation length.
\end{abstract}

\maketitle

\section{Introduction}

According to our present understanding, the Casimir effect is a
phenomenon common to all systems characterized by fluctuating
quantities on which external boundary conditions are imposed.

The confinement of quantum mechanical vacuum fluctuations of the
electromagnetic field causes long-ranged forces between two
conducting uncharged plates which is known as the (quantum mechanical)
Casimir effect \cite{C48,C53,MT97,M01}. The corresponding force between the
plates is called the Casimir force. In this form the phenomenon
was predicted in 1948 \cite{C48} by the Dutch physicist
Hendrick Casimir.

The confinement of critical fluctuations of an order parameter
also induces long-ranged forces between the system boundaries
\cite{FG78,K94,BDT00}. This is known as the statistical-mechanical
(thermodynamic) Casimir effect. In this form the effect was
discussed by Fisher and de Gennes \cite{FG78} already in 1978.

Casimir forces arise from the influence of one portion of a
system, via fluctuations, on another portion some distance away.

The best known example of a Casimir force is the van
der Waals interaction between neutral molecules. In this case the
correlations between fluctuations are mediated by photons, i.e.,
massless excitations of the electromagnetic field.
When the system is a thermodynamic one, important examples of such
massless excitations include Goldstone bosons and order parameter
fluctuations at critical points.

The quantum mechanical Casimir effect has been experimentally
verified with impressive experimental precision
\cite{HCM00} (see also Refs.\cite{MR98} and \cite{L97}).
One uses atomic force microscope techniques and measures the force
between a metallized sphere and a plate. Since it turns out
to be very difficult to keep two plates parallel with
the required accuracy, there is only one recent experiment
\cite{BSOR02} in which the original theoretical parallel plate set-up
as studied by Casimir was used. In this
experiment it has been found that the measured Casimir force
agrees with the predicted one within $15 \%$ accuracy. It is
interesting to note that at distances of the order of 10 nm between
the plates the force produces a pressure of the order of
1 atmosphere. Therefore, the Casimir effect is considered to be
very important for the design of nanoscale devices (see, e.g.,
Ref.\cite{L02} and references therein).

In statistical mechanics the Casimir force is usually characterized by
the excess free energy coming from the {\em finite-size contributions}
to the free energy of the system. The parallel plate or film geometry
turns out to be of great practical importance for experimental setups.

A useful model for the investigation of generic finite-size effects is
given by an $O(n)$-symmetric spin system ($n\geq 1$) confined to a film
geometry ($L\times \infty^2$) with periodic boundary conditions $\tau$.
Models of this sort serve as theoretical descriptions of magnets or fluids
confined between two parallel plates of infinite area. The Casimir force
per unit area in these systems is defined as
\begin{equation}
F_{Casimir}^\tau (T,L)=-\frac{\partial f^\tau_{{\rm
ex}}(T,L)}{\partial L}, \label{def}
\end{equation}
where $f^\tau _{{\rm ex}}(T,L)$ is the excess free energy
\begin{equation}
f^\tau _{{\rm ex}}(T,L)=f^\tau (T,L)-Lf_{{\rm bulk}}(T).
\label{fexd}
\end{equation}
Here $f^\tau (T,L)$ is the full free energy per unit area (and per
$k_BT$ ) of such a system and $f_{{\rm bulk}}$ is the bulk free
energy density.

According to the definition given by Eq.(\ref{def}) the
thermodynamic Casimir force is a generalized force conjugate to
the distance $L$ between the boundaries of the system with the
property $F_{{\rm Casimir}}^\tau(T,L)\rightarrow 0$ for $L\rightarrow \infty$.
We are interested in the behavior of $F_{\rm Casimir}^\tau$ when $L\gg a$,
where $a$ is a typical microscopic length scale. In this limit
finite size scaling theory is applicable. The {\em sign} of the
Casimir force is of particular interest. It is supposed that if
the boundary conditions $\tau$ are the same at both surfaces
$F_{{\rm Casimir}}^\tau$ will be {\it attractive}
\cite{es,ParryEvans,positivneg2} (strictly speaking, for an
Ising-like system this should hold above the wetting transition
temperature $T_w$ \cite{es,ParryEvans,D88}). In the case of a fluid
confined between identical walls this implies an attractive force between
the walls for large separations. When the boundary conditions {\em differ}
between the confining surfaces, the Casimir force is
expected to be {\it repulsive} \cite{es,positivneg1,positivneg2}.
The current experimental situation will be discussed later in the article.
Here we only mention that these experiments are in qualitative and in some
cases even in quantitative agreement.

In this article we discuss the behavior of the thermodynamic
Casimir force in systems with short-ranged and with subleading long-ranged
(dispersion) forces, which are present, e.g., in real fluids. Both
interactions lead to the same universality class, provided that the
dimensionality $d$ of the system and the symmetry of the ordered state are
the same. Despite this similarity we will see that, in comparison with systems
with short-ranged forces, new important finite-size contributions exist in
systems with dispersion forces. We shall also discuss proper boundary
conditions for systems with subleading long-ranged interactions and ww shall
reconsider several recent statements \cite{CD01} for the behavior of the
finite-size free energy and the Casimir force in systems with
short-ranged interactions and with dispersion forces.

From the definition in Eqs.(\ref{def}) and (\ref{fexd}) it is clear that one
needs to know the critical behavior of the free energy in a slab geometry
in order to derive the behavior of the Casimir force. Based on numerous
investigations it has turned out that the thermodynamic behavior of a system
near a second order phase transition exhibits scale invariance and universality
\cite{JJ2002,A84}. In order to set the stage for our considerations we first
recall certain bulk properties.

\subsection{Bulk systems}

Scale invariance and universality hold for the singular part
of a thermodynamic function. For later reference we quote the
decomposition into a regular and a singular part of
the free energy $f$ in units of
$k_B T_c$ and
per unit volume of, e.g., an Ising ferromagnet:
\begin{eqnarray}\label{free}
f(t,h) &=& f_{reg}(t,h) + f_{sing}(t,h)\nonumber \\
&=& f_{reg}(t,h) + |t|^{2 - \alpha} A_1
F_{\pm}\left(A_2 h |t|^{-\Delta}\right),
\end{eqnarray}
where $t = (T - T_{c})/T_{c} \gtrless 0$ is the reduced temperature, $h$ is the external
magnetic field, $\Delta$ is the critical exponent associated with the magnetic field,
$\alpha$ is the critical exponent of the specific heat, $A_1$ and $A_2$ are nonuniversal
(system dependent) metric factors, and $F_{\pm}$ are universal scaling functions.

In the scaling limit the two-point correlation function in zero field, which is of
particular interest in the present context, has the form
\cite{Fisher65,Fisher71}
\begin{equation}\label{correlf}
G(r,t\gtrless 0) = D r^{-(d - 2 + \eta)} g_{\pm}(r/\xi_{\pm})
\end{equation}
with
$\xi_{\pm}$ as the correlation length given by
\begin{equation}\label{xi}
\xi_{\pm}(t,h=0) = \xi_0^{\pm} |t|^{-\nu}, \ t \rightarrow 0.
\end{equation}
The nonuniversal constants $D$ and $\xi_{0}^{+}$ can be related to $\xi_0^{-}$,
$A_1$,  and
$A_2$
via $\xi_0^{+}/\xi_0^{-}=Q_1$, $A_1=Q_2(\xi_0^+)^{-d}$, and
$A_2=Q_3 \sqrt{D}(\xi_0^+)^{(d+2-\eta)/2}$ with $Q_1$, $Q_2$,
and $Q_3$ being universal, which leads to the
hyperuniversality hypothesis in the form of two-scale factor universality \cite{PHA91}.
In the form given above Eqs. (\ref{correlf}) and (\ref{xi}) are valid for Ising-like
systems only. For $O(n)$ models, $n\ge2$, one has in addition to take into
account that
$\xi_{-}(t\le 0) \equiv\infty$.
The universal scaling function $g_{+}(x)$ decays exponentially for $x \gtrsim 1$.
The physical origin for the onset of scale invariance can be traced back to the
divergence of the correlation length $\xi_{\pm}$. Consensus has emerged that these
statements hold in systems governed by short-range interaction potentials, i.e.,
decaying exponentially or being of a finite range.

If the exchange interaction in an Ising model on a lattice in $d$ dimensions decays
algebraically,
\begin{equation}
\label{sint}
J({\bf r}) = \frac{J}{1+(r/a)^{d+\sigma}}, \ r \equiv |{\bf r}|\ge a>0,
\end{equation}
where $a$ is the lattice constant, the value of the decay exponent
$\sigma$ is crucial with respect to universality. For
$\sigma > 2$ the leading thermodynamic critical behavior is characterized by
critical exponents
and scaling functions for short-ranged interactions
\cite{FMN72}.
Mean-field theory holds for $d>d_c=4$ irrespective of the value of $\sigma$.
For $\sigma < 2$ the  upper critical dimension is reduced to $d_c(\sigma) = 2\sigma$
\cite{FMN72,AF88}
and the values of the critical exponents depend on $\sigma$ for
$\sigma<d<d_c(\sigma)$ \cite{FMN72,BJG76,FS82}.
The crossover from short-ranged to long-ranged critical behavior occurs for
$\sigma=2-\eta_{sr}(d)$,
where $\eta_{sr}(d)$ is the critical exponent for the short-ranged system
(for a given fixed spatial dimension $d$) \cite{C96,S73,HN89,H90,J98}.
This crossover has recently been
reexamined numerically in $d = 2$ in Ref. \cite{LB2002}.

Fluids are governed by dispersion forces. In
the sense of Eq.(\ref{sint}) dispersion (van der Waals) forces in $d = 3$
dimensions are characterized by $\sigma = 3$ in the nonretarded case and
by $\sigma = 4$ in the retarded case. Therefore, the leading
thermodynamic critical
behavior of a fluid is characterized by critical exponents and scaling
functions for short-ranged interactions and the contributions due to
the power law decay of the interaction potential lead to corrections
to the asymptotic scaling behavior. Thus for $\sigma > 2$ we  refer to
interaction potentials governed by Eq.(\ref{sint}) as
subleading long-ranged interactions.

\begin{figure}
\includegraphics[width=\columnwidth]{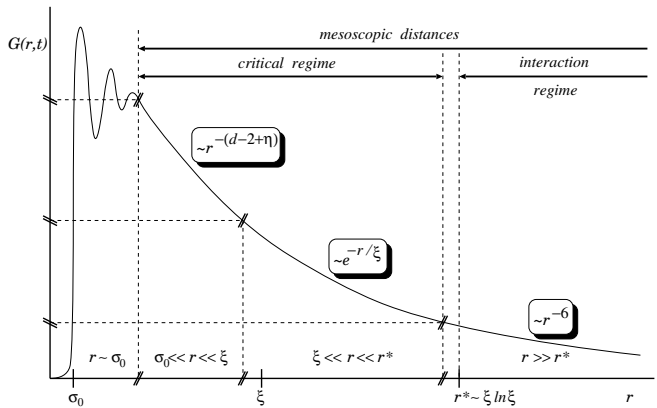}
\caption{Schematic view of the density-density correlation function $G(r,t)$
in a fluid governed by dispersion forces in $d=3$. The
behavior is shown on various length scales, where the tilted double slashes
$//$ indicate breaks in scale. For microscopic distances of the order of the
particle diameter $\sigma_0$ ``packing'' effects lead to oscillations
decaying exponentially \protect\cite{CEHH94}. Beyond  a crossover regime (not shown)
the correlation function decays according to the power law given in
Eq.(\protect\ref{correlf}) as long as $r \ll \xi$. Beyond another crossover
(not shown) the further decay is exponential for $r \gg \xi$ which finally
crosses over to the interaction dominated regime for $r > r^*$, where the
ultimate decay of the correlation function follows the decay of the interaction
potential $\sim \ r^{-6}$ in the nonretarded regime and $\sim \ r^{-7}$ in
the retarded regime (not shown). The behavior $r^* \sim \xi \ln \xi$ of the
crossover distance $r^*$ \protect\cite{KR84,FD95,D01} illustrates that  $r^*$
diverges more strongly than the correlation length $\xi$ in the vicinity of
$T_c$, i.e., the critical regime, in which the universal properties hold,
expands as $T$ approaches $T_c$.
\label{corplot}}
\end{figure}

As an illustration of this case Fig. \ref{corplot} displays schematically the
two-point correlation function  $G(r,t)$.
The universal decay of $G(r,t)$  is  governed
by Eq.(\ref{correlf}) only within the critical regime. For distances $r$
smaller than the lower limit of this regime generically nonuniversal
microscopic effects govern the behavior of the correlation function. For
distances larger than the upper limit of this regime the interaction
potential itself governs the further decay of the correlation function.
Note that $r^*/\xi \to \infty$ as $T \to T_c$, i.e., the critical regime
expands as the critical point is approached \cite{KR84,FD95,D01}. The widening of
the critical  regime leads to the divergence of
the compressibility $\kappa(t) \sim \int_0^\infty dr r^2 G(r,t)\sim |t|^{-\gamma}$, for
$t\rightarrow 0$.

For short-ranged interactions one has to bear in mind that
Eq.(\ref{correlf}) is also only valid within a critical regime $r < r^*_{\rm sr}$.
Exact results for the two-dimensional Ising model \cite{CW73}
and mean-field results
\cite{FB67} suggest that $r^*_{{\rm sr}}\sim \xi^2$, whereas
 the spherical model yields the estimate $r^*_{{\rm sr}}\sim \xi^3$ \ \cite{CD2000}.
For $r\gg r^*_{{\rm sr}}$ the correlation function decays again exponentially,
but it contains a nonuniversal prefactor \cite{CD2000}, i.e., the {\it
leading-order} behavior becomes nonuniversal. This demonstrates that in the case of
short-ranged interactions the width of the critical regime is much larger
than for a corresponding system with subleading long-ranged interactions.

\subsection{Finite size scaling}

Finite-size scaling asserts that near the bulk critical temperature $T_c$ the
influence of a finite sample size $L$ on critical phenomena is governed by
universal finite-size scaling functions which depend on the ratio $L/\xi$, so that the
 rounding of the thermodynamic singularities sets in for $L/\xi\simeq O(1)$
\cite{Fisher72,fisherbarber72,barber83,PF84,privman90,BDT00}.

From the above discussion of the behavior of $G(r,t)$ one expects
that deviations from standard finite-size scaling behavior will be
observed for $L \gg r^*$, where $r^*$ is a crossover length with
the property $r^* \gg \xi$. In particular one has $r^* \sim \xi^2$
for the Ising model or within mean-field theory, $r^* \sim \xi^3$
for the spherical model \cite{CD2000}, and $r^* \sim \xi\log\xi$
for subleading long-ranged interactions \cite{D01,DR01}.

This view has recently been challenged by Chen and Dohm \cite{CD01} who purportedly
report, for systems with short-ranged  interactions, leading finite-size contributions
different from the ones expected from the above discussion. This would
invalidate the  current understanding of finite-size scaling.
In particular these results can lead to the expectation of a
{\em nonuniversal} Casimir force at $T_c$ for a fluid between parallel
plates at a distance $L$. If correct this would be of major theoretical
\cite{KD,K94,K99} and experimental interest \cite{ML99,GC99,GC02}.
Specifically, based on exact results for the $O(n)$ symmetric $\phi^4$
field theory in the large-$n$ limit (mean spherical model),
the authors of Ref. \cite{CD01} report  the following result
for the singular part $f_{s}(t,L,\Lambda)$ of the finite-size
contribution $f(t,L,\Lambda)$ to the free energy density of a system with
periodic boundary conditions and purported short-ranged interactions in $2 < d < 4$:
\begin{equation} \label{CDsr}
f_s(t,L,\Lambda)=L^{-2} \Lambda^{d-2} \Phi(\xi^{-1}\Lambda^{-1})
 + L^{-d}X^{sr}(L/\xi).
\end{equation}
The parameter $\Lambda$ is an ultraviolet momentum cutoff and the function $\Phi$
has the property
$\Phi(0) > 0$. For the Casimir force defined by
\begin{equation} \label{CDCasimir}
F_{Casimir}(t,L,\Lambda) \equiv
-\frac{\partial}{\partial L} \{L [f(t,L,\Lambda)-f(t,\infty,\Lambda)] \},
\end{equation}
Eq.(\ref{CDsr}) implies a leading nonuniversal (cutoff dependent)
non-scaling term $\sim L^{-2}$ in the behavior of the Casimir force, because the
scaling function
$X^{sr}(x)\sim \exp(-x)$ when $x \gg 1$
\cite{Fisher72,fisherbarber72,barber83,privman90,BDT00}.
Therefore Eqs.(\ref{CDsr}) and (\ref{CDCasimir}) would also imply {\em
nonuniversal} Casimir amplitudes
\begin{equation}
\Delta_{Casimir}(d)= \Lambda^{d-2} \Phi(0)
\end{equation}
in $d > 2$ dimensions.

In the following we shall show that the results reported in Ref. \cite{CD01} can
be traced back to using a peculiar model in which the interactions are
neither  short-ranged nor of the subleading long-ranged type so that the
model does not relate to any physical realization. We find that if the
periodicity and analyticity of the Fourier transform $J({\bf k})$ of the
interaction $J(r)$ at the boundary of the Brillouin zone (in the case of
a lattice model) and the analyticity of $J({\bf k})$ at the cutoff $k=\Lambda$
(in the case of an off-lattice model) are preserved in the theoretical
analysis, then the $L^{-2}$ term in Eq.(\ref{CDsr}) {\it vanishes
identically}. We  also show that the presence or absence of this term
does {\em not} depend on the range of the interactions. If the above
 requirements for $J({\bf k})$ at the boundary of the Brillouin zone or at
$k = \Lambda$ are violated, a corresponding non-universal, non-scaling term
of order $L^{-2}$ will be observed in the finite-size behavior of
{\em any} thermodynamic function. A discussion on the influence of the cutoff on
the finite-size behavior of the susceptibility has already been presented in Ref. \cite{DR01};
see also the ``note added in proof'' of  Ref. \cite{CD}.
In Sections II and III we
present a general and unified approach that is designed to
 avoid  similar artificial effects; this should be useful also in the context of
quantum phase
transitions and field theory.  In Section III
we also  summarize the present state of knowledge on the finite-size behavior of
systems with subleading long-range interactions, focusing on the expected
behavior of the singular part of the free energy. One should distinguish between
the  cases {\it i)} $d+\sigma<6$ and {\it ii)} $d+\sigma=6$, which
contains the physically most important case of dispersion forces in $d=3$ with
$\sigma=3$.
In case {\it ii)} one expects additional
logarithmic finite-size contributions. The article closes in Sect. IV with a summary and
concluding remarks, where we also discuss possible boundary
conditions for systems with subleading long-ranged interactions.

\section{Finite-size behavior of the free energy density}

In this section we present our critique of the finite-size scaling analysis
of the free energy and the Casimir force presented in Ref. \cite{CD01}. As
pointed out already the statements of Ref. \cite{CD01} are based on exact results for
the $O(n)$ symmetric $\phi^4$ field theory in the large-$n$ limit (mean spherical model)
with periodic boundary conditions.

\subsection{Analytical properties}

Before we turn to the finite-size analysis, we discuss  briefly the
consequences of the assumptions for the Fourier transform of the
interaction $J({\bf k})$ used in Ref. \cite{CD01} for the {\it bulk}
properties of the model. For a system on a lattice $J({\bf k})=
\sum_{\bf r}J({\bf r})\exp(i{\bf k}.{\bf r})$, where the sum runs over the  lattice sites.
For an off-lattice system the sum has to be replaced by the corresponding integral.

\begin{figure}[btf]
\includegraphics[width=\columnwidth]{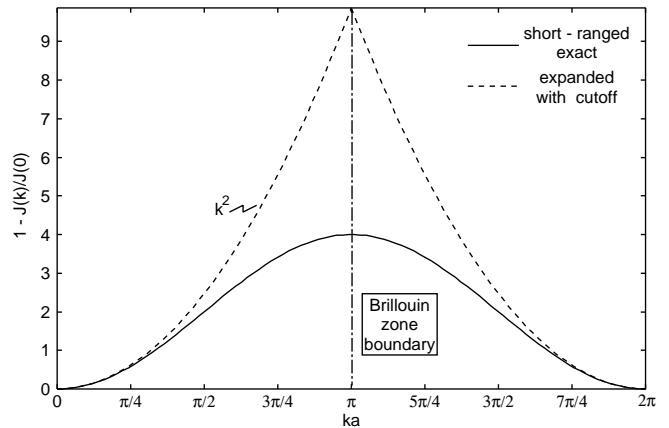}
\caption{Dispersion relation $1 - J(k)/J(0)$ as function of $k$ for
nearest-neighbor interactions in one dimension (solid line) in comparison
with a $k^{2}$-spectrum with a sharp cutoff at the Brillouin zone boundary,
for $d=1$ dimension, as implemented in Ref. \protect\cite{CD01} (dashed line).
The zone boundary is marked by the vertical dash-dotted line and $a$ is the
lattice constant. Note that the application of a sharp cutoff to a pure
$k^{2}$ spectrum in the first Brillouin zone implies an artificial cusp-like
nonanalyticity of the dispersion relation at the zone boundaries.  This is
absent for genuine short-ranged interactions. \label{Jofkplot}}
\end{figure}

In Fig. \ref{Jofkplot} we compare $J({\bf k})$ for
a short ranged (nearest neighbor) lattice model (see, c.f., Eq.(\ref{Jofk}))
with the standard $k^{2}$ spectrum in the infrared limit for short-ranged
interactions in the spirit of Ref. \cite{CD01} endowed with a cutoff
$\Lambda = \pi/a$ at the boundaries $\pm \pi/a$ of the first Brillouin zone
in one dimension. The restriction to one dimension only simplifies the
notation and is not essential for the following arguments.
>From Fig. \ref{Jofkplot} it is obvious that the continuation of the
$k^{2}$-spectrum, which is  correct only in the infrared limit
$k\rightarrow 0$, to the full Brillouin zone and its truncation at the zone
boundaries introduces a cusp-like nonanalyticity into the spectrum. This
nonanalyticity is artificial and is not a generic feature of short-ranged
interactions. As we shall demonstrate below, this artificial choice is the
reason for the non-scaling and non-universal finite-size effects $\sim L^{-2}$
reported in Ref. \cite{CD01}. In fact, the importance of the properties of the
dispersion relation at the Brillouin zone boundary for the critical finite-size
behavior of the free energy in $d=2$ dimensions has already been mentioned
by Cardy (see Eq.(3.12) of Ref. \cite{C86}) in the course of deriving
the Casimir amplitude $\Delta_{Casimir}(2) = -\pi c/6$ in $d=2$ for periodic
boundary conditions, where $c$ is the central charge of the model under
consideration (e.g., $c = 1/2$ for the critical $2d$ Ising model with short
ranged interacions).

It is instructive to investigate the consequences of a nonanalytic spectrum
of the kind displayed in Fig. \ref{Jofkplot} in real space. According
to Ref. \cite{CD01} the corresponding correlation function reads for $r\gg\xi$
in $d$ dimensions
\begin{eqnarray}\label{cCD}
G(r,t) &=& 2\Lambda^{d-2} (2 \pi r \Lambda)^{-(d+1)/2}
\frac{\sin[\Lambda r - \pi (d-1)/4]}{1+\xi^{-2}\Lambda^{-2}}\nonumber \\
&+& O(\exp(-r/\xi)).
\end{eqnarray}
Thus the correlation function decays according to a power law with the decay
exponent $(d+1)/2$
rather than  exponentially. Furthermore,
the correlation function oscillates with a period set by the inverse of the
cutoff $\Lambda$. For separations $r > r^{*}$ (see
Fig. \ref{corplot}) Eq.(\ref{cCD}) therefore implies that the interaction
potential for a system with a truncated $k^{2}$-spectrum as shown in
Fig. \ref{Jofkplot}, should be {\em leading long-ranged}   rather than
{\em short-ranged}  or even {\em subleading long-ranged} in
$2 \leq d \leq 4$, but also containing both positive and negative  contributions.
Therefore, for all spatial dimensions of physical relevance, the model
investigated in Ref. \cite{CD01} would appear to correspond to a model with competing
leading long-ranged interactions in real space. For such a model the
finite-size scaling as developed for systems with short-ranged or subleading
long-ranged interactions is not expected to be applicable.

\subsection{Finite-size properties of the $O(n\rightarrow\infty)$ model}

We substantiate our view by turning to a detailed analysis of the
finite-size behavior of the
free energy and other thermodynamic functions, i.e., we provide an account
of the mathematical mechanism that produces the non-universal
and non-scaling leading finite-size effects reported in Ref. \cite{CD01}.

As a case study we quote the expression for the total free energy density
of a fully finite mean spherical model
(the $n\rightarrow\infty$ limit of an $O(n)$ model)
with nearest-neighbor interactions of strength $J$ on a hypercubic lattice,
which is given by \cite{BDT00}
\begin{eqnarray}\label{freesphe}
&&\beta f(K,h|{\bf L}, {\bf \Lambda}) = \frac{1}{2}\sup_{\tau>0}\left\{
-\frac{h^2}{K\tau}+\frac{1}{L_1 \times \cdots \times L_d}\right.\nonumber \\
&&\quad \left. \times \sum_{\bf k}\ln\left[\tau+\sum_{j=1}^d
2(1-\cos k_j)\right] - K\tau\right\} \\
&&\  +\frac{1}{2}\left[\ln\frac{K}{2\pi}-2dK\right],\nonumber
\end{eqnarray}
where $K=\beta J$ and $h$ is a properly normalized magnetic field. The
parameter $\tau$ is related to the correlation length $\xi$ via
$\tau = \xi^{-2}$ \cite{rem1} which relates $\tau$ also to the reduced temperature $t$.
The relevant physical information of Eq.(\ref{freesphe}) is contained in the
sum over the wave vectors ${\bf k} = (k_1,\dots,k_d)$ in the first Brillouin
zone of the simple
cubic lattice. For general interaction potentials on general lattices
this sum involves the dispersion relation $\omega({\bf k})$, where, e.g.,
$\omega({\bf k}) = \sum_{j=1}^d 2 (1 - \cos k_j)$ for nearest-neighbor
interactions on a simple cubic lattice (see Eq.(\ref{freesphe}) and, c.f.,
Eq.(\ref{Jofk})). In order to provide a general
description of finite-size scaling of the free energy, we will
therefore consider the quantity
\cite{BDT00} (see also Eq.(14) in Ref. \cite{CD01})
\begin{equation}\label{udef}
U_{d,\sigma}(\tau,{\bf L}, {\bf \Lambda}) =
\frac{1}{2 L_1 \times \cdots \times L_d}\sum_{{\bf k}\in {\cal B}}
\ln [\tau + \omega({\bf k})]
\end{equation}
as function of the system size ${\bf L}$,
where ${\bf L} = (L_1,\cdots,L_d)$ denotes the set of lengths that determine
the geometry of the system, ${\bf \Lambda} = (\Lambda_1,\cdots,\Lambda_d)$
is the set of cutoffs in reciprocal (i.e., $k$) space and $\tau$ is a quantity
proportional to the reduced temperature $t$, e.g., $\tau = \xi^{-2}$  for the mean
spherical model with short-range interactions. For lattice systems ${\bf k}$ belongs to
the first Brillouin zone ${\cal B}$ and if the system is on a hypercubic lattice one has
$\Lambda_i=\pi/a_i$, where $a_i$ is the lattice spacing along the direction
$i$, $i=1, \cdots, d$.  For off-lattice systems the summation is carried out
over those values ${\bf k}\in {\cal B}$ that fulfill the requirements
$-\Lambda_i\le k_i<\Lambda_i$, $i=1,\cdots,d$. However, for off-lattice systems,
 there is no obvious choice for the cutoff. One usually takes $\Lambda
\simeq \tilde{a}^{-1}$, where $\tilde{a}$ is some fixed characteristic microscopic length of the
system. For a fluid $\tilde{a}$ can be taken to be the diameter $\sigma_0$ of
a fluid particle (see also Fig. \ref{corplot}).
In Eq. (\ref{udef}) the subscript $\sigma$ characterizes the range of
the interaction. The fluctuation spectrum $\omega({\bf k})$ of the
order parameter is given as a linear function of the
Fourier transform $J({\bf k})$ of the interaction (see, c.f., Eqs.(\ref{Jofk}) and
(\ref{wsr})).

The formal expression given by Eq. (\ref{udef}), on which our specific considerations in this
subsection
are based,  has widespread applications. The
expression in Eq. (\ref{udef})  always appears as the one-loop contribution
to the free energy in field-theoretic Ginzburg-Landau models \cite{KD}.
The line of arguments presented here is therefore  of general importance
and is not limited to  the specific model under consideration here.

For periodic boundary conditions $L_i \Lambda_i/(2\pi)\equiv
M_i$ are integer numbers, $i=1,\cdots,d$, where
$\prod_{i=1,\cdots,d}M_i\equiv N$ fixes the number of degrees of
freedom in the system. The values of the components $k_i$ of the
vector ${\bf k}$ are given by $k_i=2\pi m_i/L_i$, with $-M_i\le m_i\le M_i-1$,
$i=1,\cdots,d$.

In order to analyze the sum in Eq.(\ref{udef}) we use the Poisson summation formula
\begin{equation}\label{P}
    \sum_{m=a}^{b} f(m)=\sum_{n=-\infty}^{\infty} \int_{a}^{b} dm \
    e^{i \; 2\pi m n} f(m) +\frac{1}{2}\left(f(a)+f(b)\right).
\end{equation}
After some algebra one obtains
\begin{equation}\label{ud}
U_{d,\sigma}(\tau,{\bf L}, {\bf \Lambda})=U_{d,\sigma}(\tau, {\bf
\Lambda})+ \Delta U_{d,\sigma}(\tau,{\bf L}, {\bf \Lambda}),
\end{equation}
where
\begin{equation}\label{udb}
U_{d,\sigma}(\tau, {\bf
\Lambda})=\frac{1}{2(2\pi)^d}\int_{-\Lambda_1}^{\Lambda_1}dm_1
\cdots \int_{-\Lambda_d}^{\Lambda_d}dm_d \ \ln[\tau+\omega({\bf m})]
\end{equation}
takes into account the contributions of the bulk system, while
\begin{eqnarray} \label{udd}
&&\Delta U_{d,\sigma}(\tau,{\bf L}, {\bf \Lambda})=\frac{1}{2(2\pi)^d}
\sum_{{\bf n}\ne {\bf 0}}\int_{-\Lambda_1}^{\Lambda_1}dm_1 \cdots
\int_{-\Lambda_d}^{\Lambda_d}dm_d \nonumber \\
&& \ \times \exp{\left(i \sum_{j=1}^d n_j m_j L_j \right)}
\ln[\tau+\omega({\bf m})]
\end{eqnarray}
incorporates all contributions due to the finite size of the system. For
further analysis of the Casimir effect  the film geometry $L\times\infty^{d-1}$
is the most relevant one.  It is obtained from Eq.(\ref{udd}) in the limit
$L_2\rightarrow\infty,\cdots,L_d\rightarrow\infty$, setting
$L_1 \equiv L$. In order to simplify the notation we finally set
$\Lambda_1 = \Lambda_2 = \cdots = \Lambda_d \equiv \Lambda$ so that
\begin{eqnarray} \label{uddfilm}
&& \Delta U_{d,\sigma}(\tau,L,\Lambda)=
\frac{1}{2(2\pi)^d}\sum_{n_1\ne 0}\int_{-\Lambda}^{\Lambda}dm_1
\cdots \int_{-\Lambda}^{\Lambda}dm_d \nonumber \\
&& \ \times \exp{\left(i n_1 m_1 L \right)}\ln[\tau+\omega({\bf m})],
\end{eqnarray}
which after two integrations by parts with respect to $m_1$
can be rewritten as
\begin{eqnarray} \label{ufinal}
&&\Delta U_{d,\sigma}(\tau,L,\Lambda)=\Delta U_{d,\sigma}^{(1)}(\tau,L,\Lambda)
 \nonumber \\
&&\ -L^{-2} \frac{1}{2(2\pi)^d}\sum_{n_1\ne
0}\frac{1}{n_1^2}\int_{-\Lambda}^{\Lambda}dm_1 \cdots
\int_{-\Lambda}^{\Lambda}dm_d \nonumber \\
&&\  \times \exp{\left(i n_1 m_1 L \right)} \
\partial_{m_1}\left[\frac{\partial_{m_1}\omega({\bf
m})}{\tau+\omega(\Lambda, m_2,\cdots,m_d)}\right],
\end{eqnarray}
with
\begin{eqnarray}
\label{CDterm}
&& \Delta U_{d,\sigma}^{(1)}(\tau,L,\Lambda)=L^{-2}\frac{\pi^2}{6}
\frac{1}{(2\pi)^d} \int_{-\Lambda}^{\Lambda}dm_2 \cdots
\int_{-\Lambda}^{\Lambda}dm_d \nonumber \\
&& \ \times \frac{\partial_{m_1}\omega({\bf m})|_{m_1=\Lambda} -
\partial_{m_1}\omega({\bf m})|_{m_1=-\Lambda}
}{\tau+\omega(\Lambda,m_2,\cdots,m_d)}.
\end{eqnarray}
The above expression is obtained by identity transformations of the initial
sum and is valid both for lattice and off-lattice systems. In the analysis
below we show that it is the term $\Delta U_{d,\sigma}^{(1)}(\tau,L,\Lambda)$ which
produces the contributions on which the statements of Ref. \cite{CD01} are
based.

First, we evaluate this term for lattice systems. If the interactions
$J({\bf r})\ge 0$ are such that they depend only on the distances
between the particles, and these are the only interactions we are concerned with
here, then $J({\bf k})=J(-{\bf k})$.  Since there is no physical reason for
singularities anywhere except at $k=0$ the derivatives of $J({\bf k})$ with
respect to ${\bf k}$ should exist at least for all $k\ne 0$ and therefore
$\partial_{\bf k}J({\bf k})=-\partial_{\bf k}J(-{\bf k})$. This  holds for
lattice and off-lattice systems. For lattice systems $J({\bf k})$ is a
periodic function with the property
\begin{equation}\label{wdef}
J({\bf k} + 2 \Lambda_i {\bf e}_i) = J({\bf k}),
\end{equation}
where ${\bf e}_i$ is a unit vector in  reciprocal space. This implies
that $\partial_{\bf k}J({\bf k})=0$ at the borders of the Brillouin zone
and therefore
\begin{equation}
\Delta U_{d,\sigma}^{(1)}(\tau,L,\Lambda)\equiv 0.
\end{equation}
Note that the above result does not depend on the range of the interaction -
it is true for short-ranged, subleading long-ranged, as well as for leading
long-ranged interactions. As an illustration of the above general arguments we
recall that the exact Fourier transform of the nearest neighbor interaction
on a $d$-dimensional hypercubic lattice reads
\begin{equation} \label{Jofk}
J({\bf k}) = 2J \sum_{j=1}^d \cos k_j \equiv J\left(2d-\omega({\bf k})\right),
\end{equation}
where
\begin{equation}\label{wsr}
\omega({\bf k}) \equiv \sum_{j=1}^d 2(1-\cos k_j).
\end{equation}
The general properties of $J({\bf k})$ that we have discussed above can be easily
verified from Eqs. (\ref{Jofk}) and (\ref{wsr}).

We now reconsider the quantity $\Delta U_{d,\sigma}^{(1)}(\tau,L,\Lambda)$ if the exact
spectrum is replaced by its asymptotic form, valid in the infrared limit
$k\rightarrow 0$, for all ${\bf k}\in {\cal B}$. This is a very common
procedure in the theory of critical phenomena based on the general idea that
only long wave-length (small $k$) contributions are important for the critical
properties of the system. This leads to $\omega({\bf m}) = {\bf m}^2$, which
is the spectrum used in Ref. \cite{CD01}. One immediately obtains
\begin{eqnarray} \label{UCDSR}
\Delta U_{d,\sigma}^{(1)}(\tau,L,\Lambda) &=&
\frac{\Lambda}{L^{2}}\frac{2\pi^2}{3} \frac{1}{(2\pi)^d}
\int_{-\Lambda}^{\Lambda}dm_2 \cdots \int_{-\Lambda}^{\Lambda}dm_d\nonumber \\
&&\times \frac{1}{\tau + \Lambda^2+m_2^2+\cdots+m_d^2} \nonumber \\
&=& \frac{\Lambda^{d-2}}{L^{2}}\frac{1}{6}
\frac{1}{(2\pi)^{d-2}} \int_{-1}^{1}dm_2 \cdots \int_{-1}^{1}dm_d\nonumber \\
&& \times \frac{1}{1 + \tau /\Lambda^2+m_2^2+\cdots+m_d^2}.
\end{eqnarray}
We recall that in the spherical limit of the $O(n)$ model one has $\tau = \xi^{-2}$.
 Eq.(\ref{UCDSR}) {\em exactly} reproduces the nonuniversal
leading but non-scaling finite-size contribution to the free energy as
reported in Eq.(16) of Ref.\cite{CD01} for the corresponding field-theoretical
model. Similar contributions exist also for subleading long-ranged
interactions to which we turn in Sect.III. According to the above
considerations such nonuniversal (cutoff dependent) contributions of the
order of $L^{-2}$ will always appear if
\begin{equation}
\label{cond}
\left.\frac{\partial J({\bf k})}{\partial{k_1}}\right|_{k_1=\Lambda}
\ne \left.\frac{\partial J({\bf k})}{\partial{k_1}}\right|_{k_1=-\Lambda}.
\end{equation}
Note that only the properties of the Fourier transform $J({\bf k})$ of the
interaction at the boundary of the set ${\cal B}$ of allowed $k$ values is
important here. For a field-theoretic model these are definitely a matter of
definition. For lattice models these properties follow automatically.
Approximating the spectrum $\omega({\bf k})$ by its infrared asymptotic
behavior leads to an artificial cusp-like singularity at the border of the
Brillouin zone as is illustrated in Fig.~{\ref{Jofkplot}}. A corresponding
approximation for Eq.(3.11) in Ref. \cite{C86} would lead to an incorrect
prediction of the critical finite-size contribution to the free energy
(see Eq.(3.12) in Ref.\cite{C86}) leading to a vanishing Casimir amplitude.

Before we consider how to modify the definition of the continuous
field-theoretic model as to avoid a nonzero
$\Delta U_{d,\sigma}^{(1)}(\tau,L,\Lambda)$, we make some general remarks.
First, the considerations presented above can  easily be extended to any
geometry of the type $L^{d-d'}\times \infty^{d'}$, $0\le d'\le d$.
Second,  in the
above discussion we did {\em not} specify the type of the interactions --
short-ranged, leading long-ranged, or subleading long-ranged. This implies
that the $L^{-2}$ corrections in question
exist for {\em any} type of interaction for periodic boundary conditions,
provided Eq.(\ref{cond}) is valid. Furthermore,
further integrations by parts yield additional  contributions of the order
$L^{-4}$, $L^{-6}$, etc. In $2<d<4$ dimensions only the term $L^{-2}$ is
important, but in $d>4$ spurious finite-size terms of the orders $L^{-2}$
and $L^{-4}$ will be generated. Finally, we note that the precise form of
the integrand in Eq. (\ref{uddfilm})  was not used in the above analysis.
This implies
that spurious $L^{-2}$ corrections also occur in other quantities such as
the susceptibility \cite{CD99,CD} (see also the discussion in Appendix G
of Ref.\cite{DR01}), the specific heat \cite{CDcm},
etc. The influence of different cutoff types (sharp or smooth) and of the
truncation of the expansion of the Fourier transform of the interaction on
the finite-size behavior of the susceptibility has been considered in
detail in Ref.\cite{DR01}. The only difference with respect to the free energy
is that the susceptibility diverges as $L^{\gamma/\nu}$ at $T_{c}$
leaving any $L^{-2}$ contribution as a {\em small} correction, whereas
the singular part of the free energy behaves as $L^{-d}$ and therefore the
cutoff dependent term of the order $L^{-2}$ becomes {\em dominant} in the
critical region for $d > 2$.

Since there is no physical reason for the introduction of a
{\em sharp} cutoff in the $k$-space representation of a field-theoretic model,
one option to avoid artificial $L^{-2}$ contributions is to implement
of a {\em smooth} cutoff. Various forms of smooth cutoffs are possible
\cite{DR01,CD}. For example in Ref.\cite{CD} a modified continuum
Ginzburg-Landau Hamiltonian has been considered (see also Ref.\cite{P92}):
\begin{equation}
\label{Hmod}
H=\int_{V}d^d x \left[ \frac{1}{2}r_0\varphi^2+\frac{1}{2}(\nabla \varphi)^2+u_0 (\varphi^2)^2
+\frac{1}{2\Lambda^2}(\nabla^2\varphi)^2\right].
\end{equation}
The last term in Eq.(\ref{Hmod}) introduces the smooth cutoff which is
parameterized by a wave number $\Lambda$. The finite-size effects of
thermodynamic quantities differ substantially for the above Hamiltonian
and the standard one with a sharp cutoff: In the framework corresponding to
Eq.(\ref{Hmod}) thermodynamic quantities approach their bulk value {\em
exponentially} as function of $L$ for $T \neq T_c$ fixed, whereas for the
standard Ginzburg - Landau Hamiltonian with a sharp cutoff the bulk limit
is reached according to the power law $\sim L^{-2}$ for any temperature.
In particular, this has been observed for the
susceptibility \cite{rem2}, the specific heat \cite{CDcm}, and the free
energy \cite{CD01}. The effect of a smooth cutoff prodecure on the finite-size
behavior of the susceptibility was also discussed in detail in Ref.
\cite{DR01}, where the aforementioned exponential finite-size behavior
was recovered. As already noticed in Refs. \cite{CD2000}, \cite{CD01},
and \cite{CD} the presence of a sharp cutoff is mandatory for the occurrence
of the aforementioned nonuniversal $L^{-2}$ contributions to finite-size
scaling. For general $d>2$ it was also realized that a close relationship
exists between a non-exponential large-distance behavior of the bulk
correlation function (generated by the sharp cutoff) and the power-law
finite-size behavior of both the susceptibility above $T_c$ \cite{CD2000}
and the singular part of the free energy \cite{CD01}. Here we have
demonstrated that all such finite-size effects arise from a single origin
and are unphysical mathematical artifacts due to the imposed singularity of
$J(k)$ at the boundary of the allowed $k$-values. This, in turn, generates
long-ranged correlations in real space. In order to eliminate spurious
finite-size contributions we propose the replacement (see Eq.(\ref{uddfilm}))
\begin{eqnarray}
\label{repl}
&&\Delta U_{d,\sigma}(\tau,L,\Lambda) \to \Delta U_{d,\sigma}(\tau,L,\Lambda)
-\Delta U_{d,\sigma}^{(1)}(\tau,L,\Lambda) \nonumber \\
&&\ =\frac{1}{2(2\pi)^d}\sum_{n_1\ne
0}\int_{-\Lambda}^{\Lambda}dm_1 \cdots \int_{-\Lambda}^{\Lambda}dm_d\nonumber\\
&&\quad \times \exp{\left(i n_1 m_1 L \right)}\ \ln[\tau+\omega({\bf m})] \\
&&\ - L^{-2}\frac{\pi^2}{6}
\frac{1}{(2\pi)^d}\int_{-\Lambda}^{\Lambda}dm_2 \cdots
\int_{-\Lambda}^{\Lambda}dm_d\nonumber \\
&&\quad \times \frac{\partial_{m_1}\omega({\bf
m})|_{m_1=\Lambda}-  \partial_{m_1}\omega({\bf m})|_{m_1=-\Lambda}
}{\tau+\omega(\Lambda,m_2,\cdots,m_d)} \nonumber
\end{eqnarray}
and corresponding replacements generated by derivatives of Eq.(\ref{repl})
with respect to the parameter $\tau$ in the definition of each model system
{\em regardless} of the implementation of a sharp cutoff. Within such a
scheme the well established methods for field-theoretic calculations in the
presence of sharp cutoff are preserved. For lattice system this is an identity
transformation, because $\Delta U_{d,\sigma}^{(1)}(\tau,L,\Lambda) \equiv 0$
as expounded above. Note that the replacements given by Eq.(\ref{repl}) and
its derivatives with respect to $\tau$ do not interfere with the treatment
of bulk systems, since $\Delta U_{d,\sigma}^{(1)}(\tau,L=\infty,\Lambda) = 0$.
They only become important for studies of the finite-size scaling behavior of
systems endowed with a sharp cutoff. These replacements remove the artificial
cutoff dependent finite-size contributions to thermodynamic quantities
in the spherical limit $n \rightarrow \infty$ of $O(n)$ models and to
one-loop order for $O(n)$ models with finite $n$.

\section{Systems with subleading long-ranged interactions}

First, we briefly recall the finite-size behavior of
systems with subleading long-ranged interactions.

In Ref. \cite{DR01} and in Refs. \cite{D01} and \cite{HD02} it was
shown that the susceptibility of a finite system with dispersion
interactions, which decay as $r^{-d-\sigma}$ for large distances,
can be written for $2<d<4$, $2<\sigma<4$, and $d+\sigma<6$ in the form
\begin{eqnarray}
\label{chi}
\chi(t,L)&=&L^{\gamma/\nu}X_\chi(L/\xi,bL^{2-\sigma-\eta}) \\
&\simeq&
L^{\gamma/\nu}[X_\chi^{sr}(L/\xi)+bL^{2-\sigma-\eta}X_\chi^{lr}(L/\xi)],\nonumber
\end{eqnarray}
where
\begin{equation}X_\chi^{sr}(x\rightarrow+\infty)\simeq X_\chi^{sr,+}x^{-\gamma/\nu}+
O(\exp(-const.\ x)).
\end{equation}
For the long-ranged part one has
\begin{equation}
X_\chi^{lr}(x\rightarrow+\infty)\simeq X_{\chi,1}^{lr} x^{-2\gamma/\nu+\sigma}+
X_{\chi,2}^{lr} x^{-2\gamma/\nu-d}.
\end{equation}
The amplitude $b$ is a
nonuniversal parameter that can be determined from the Fourier transform of
the interaction. The first term of the asymptotic behavior of
$X_\chi^{lr}(x)$ yields the bulk corrections to scaling as
predicted by Kayser and Ravech\'{e} \cite{KR84}, while the second term
yields the leading finite size correction to the susceptibility for
$L/\xi \gg 1$. This second term leads to
$\chi(t,L)-\chi(t,\infty)\sim t^{-d\nu-2\gamma}
L^{-(d+\sigma)}$, $L/\xi\gg1$, i.e., the finite-size corrections to the
bulk behavior are governed by a {\em power law} rather than an exponential
 function of the system size $L$. Finally, we note that for the
physically most important case $d+\sigma=6$ (e.g., $d = \sigma = 3$
for non-retarded van der Waals forces in $d = 3$) one finds additional
logarithmic corrections in Eq.(\ref{chi}) \cite{D01,DR01}, which can be
incorporated by the replacement $X_{\chi}^{lr}(x)$ $\rightarrow$
$X_{\chi}^{lr,1}(x)\ln L+X_{\chi}^{lr,2}(x)$ \cite{D01}.

The behavior of the susceptibility outlined above is consistent with the
behavior of the bulk pair correlation function in systems with dispersion
forces
\cite{D01}
\begin{equation}
G(r,t)=r^{-(d-2+\eta)}[g^{sr}_{\pm}(r/\xi)+r^{-(\sigma-2+\eta)}
g^{lr}_{\pm}(r/\xi)].
\end{equation}
The modified Fisher-Privman \cite{FP84} finite-size scaling hypothesis for the free
energy density in such systems can be cast into the form
\begin{eqnarray}
\label{feslr}
f_s(t,L)&=&L^{-d} X(L/\xi,bL^{2-\sigma-\eta}) \\
&\simeq& L^{-d}[X^{sr}(L/\xi) + bL^{2-\sigma-\eta}X^{lr}(L/\xi)], \nonumber
\end{eqnarray}
where $X^{sr}(x \rightarrow+\infty) \simeq X^{sr,+}x^d+O(\exp(-const. \ x))$
is the short-ranged
contribution. For the long-ranged contribution one expects
$X^{lr}(x)\simeq X_1^{lr}x^{d+\sigma+\eta-2} + X_2^{lr}x^{\eta-2}$.
The first term in the asymptotic behavior of $X^{lr}$ yields a
{\it new bulk correction to scaling} that is due to the subleading part
of the interaction (analogous to the corresponding terms predicted by
Kayser and Ravech\'{e} for the susceptibility \cite{KR84} and observed in
spherical model calculations \cite{DR01}).
Its temperature dependence for $T>T_c$ is given by
$t^{d\nu+(\sigma+\eta-2)\nu}$. For $L/\xi\gg 1$ the second term leads to
a finite-size contribution of the form $L^{-(d+\sigma)}$. As for the
finite-size scaling behavior of the susceptibility, additional logarithmic
corrections have to be added to $X^{lr}$ for $d+\sigma=6$.
Finally, we note that there may also be a third {\em constant} contribution
to the asymptotic behavior of $X^{lr}(x)$ for $x \gg 1$, which does not appear
in the susceptibility, and which leads to a $L^{-4}\ln L$ finite-size
contribution to the free energy for $d=\sigma=3$.

Eq. (\ref{feslr}) has been derived in Ref. \cite{CD01} for the case $\eta=0$
where the main focus is set on the discussion of finite-size contributions.
Furthermore, the results reported in Ref. \cite{CD01} apparently apply only for
$d+\sigma<6$, because no logarithmic corrections were found.
Thus a complete verification of Eq. (\ref{feslr}) is still missing.
We now turn to the investigation of some of the consequences of the
assumptions used in the model Ref. \cite{CD01}.

We suppose that the interaction potential $J({\bf r})$ is of the dispersion
type as defined by Eq. (\ref{sint}).
The Fourier transform of such an interaction is
\begin{eqnarray}
J({\bf k}) &\simeq& J({\bf 0}) \left(1-v_2 k^2+v_\sigma k^\sigma-v_4 k^4+
O(k^6)\right)\nonumber \\
&\equiv& J({\bf 0})-K \omega({\bf k})/\beta,
\label{int2}
\end{eqnarray}
where $k=|{\bf k}|$, $4>\sigma>2$ and $J({\bf 0})$,
$v_2$, $v_\sigma$, and $v_4$
are nonuniversal positive constants. The constant $J({\bf 0})$ is the
 ground state energy of the system and
$\omega({\bf k})\simeq  k^2-b k^\sigma+ c k^4 + O(k^6)$, where
$K=\beta v_2 J({\bf 0})$, $b=v_\sigma/v_2>0$
and $c=v_4/v_2>0$ are nonuniversal constants. Since $J({\bf r})\ge 0$ and
thus  $J({\bf 0})> J({\bf k})$ for ${\bf k}\ne {\bf 0}$, the values of
$b$ and $c$ are such, that there are no real roots of the equation
$1-bk^{\sigma-2}+ck^2=0$ with respect to $k$.

The free energy of an $O(n)$ model with an interaction described by
Eq.(\ref{sint}), in the limit $n\rightarrow\infty$, is given by the expression
\begin{eqnarray}
&&\beta f(K,h|{\bf L},{\bf \Lambda})=\frac{1}{2}\sup_{\tau>0} \left\{
-\frac{h^2}{K\tau}+\frac{1}{L_1\times \cdots \times L_d} \right.\nonumber \\
&&\left. \ \times \sum_{\bf k} \ln[\tau+\omega({\bf k})]-K \tau \right\}
+\frac{1}{2}\left[ \ln\frac{K}{2\pi}-\frac{K}{v_2}\right].
\label{fed}
\end{eqnarray}
In the presence a sharp cutoff in $k$-space this leads to
\begin{eqnarray}
\lefteqn{\Delta U_{d,\sigma}^{(1)}(\tau,L,\Lambda)=
\frac{\Lambda^{d}}{L^{2}}\frac{1}{6}
\frac{1}{(2\pi)^{d-2}}
\int_{-1}^{1}dm_2 \cdots \int_{-1}^{1}dm_d} \\
&&\ \times
\frac{1-\frac{1}{2}b\sigma\Lambda^{\sigma-2}(1+\theta^2)^{\sigma/2-1}+2c^2
\Lambda^2(1+\theta^2)}{\tau +\Lambda^2(1+\theta^2)-b\Lambda^\sigma
(1+\theta^2)^{\sigma/2}+2c^2\Lambda^4(1+\theta^2)^2},  \nonumber
\end{eqnarray}
where $\theta^2=m_2^2+\cdots+m_d^2$. This term is missing in Eq.(9) in Ref.
\cite{CD01} but it is manifestly present in the case of a  sharp
cutoff. We therefore conclude that Eq.(9) of Ref. \cite{CD01} is only correct
within the smooth cutoff procedure, but not within the sharp cutoff one.
As already explained above, Eq. (9) of Ref. \cite{CD01} coincides with our
Eq. (\ref{feslr}) for systems  with $\eta=0$ and $d+\sigma<6$ for
{\it periodic} boundary conditions. In Ref.\cite{CD01} it is supposed to be
valid also for systems with {\it Dirichlet} boundary conditions. However,
Dirichlet boundary conditions are inconsistent with the long-ranged nature
of the dispersion forces (subleading long-ranged interaction), because the
``missing neighbors'' of the ordering degrees of freedom
at a surface of such a system by the nature of long-ranged
interactions generate a long-ranged surface field. We therefore conclude that
the consideration of systems with subleading long-range interactions combined
with Dirichlet boundary conditions as proposed in Ref.\cite{CD01} is of
no physical relevance. In the next section we summarize our findings and also
comment on the proper boundary conditions and the expected finite-size
behavior of systems with dispersion forces and real boundaries. We will present
arguments as to why we expect this to differ significantly from
Eq.(\ref{feslr}).

\section{Summary and concluding remarks}

It has been shown in Sects. II and III that nonanalyticities in the dispersion
relation $\omega({\bf k})$ at the momentum cutoff lead to a bulk model
which has leading and competing long-ranged interactions in real space. Thus
even in the bulk it has peculiar properties such as the two-point correlation
function given by Eq.(\ref{cCD}). For such a model finite-size scaling
developed for systems with short-ranged or subleading long-ranged
interactions does not apply from the outset. We are not aware of any physical
system governed by such a type of interaction.

In order to investigate the critical behavior of a model system by means
of an effective Ginzburg-Landau Hamiltonian in ${\bf k}$-space, one
usually expands $\omega({\bf k})$ in the (infrared) limit $k \to 0$, keeping
only the leading term(s). Divergent momentum integrals can be
regularized by the introduction of a cutoff which is motivated by the
presence of a Brillouin zone (lattice models) or finite particle sizes
(continuum models). If a sharp cutoff in momentum space is applied as
 a {\em strict} feature of an otherwise approximate (expanded)
dispersion relation, the ensuing nonanalyticity of $\omega({\bf k})$ at the
cutoff generates long-ranged correlations and finite-size effects of
the order $L^{-2}$. These effects do not occur in actual systems with
short-ranged or subleading long-ranged interactions and therefore they are
artificial. In particular, the nonuniversal long-ranged Casimir forces
reported in Ref. \cite{CD01} have no physical relevance for systems with
short - ranged forces.

It is instructive to consider the finite-size behavior of the free energy
in systems with subleading long-ranged interactions. As is well known, the
free energy  decomposes into a sum of a regular and a singular part.
In Section III
we have discussed the finite-size behavior of the singular part
for the case of periodic boundary conditions in a film geometry. However, the
regular part is also important and it has experimental consequences for the
Casimir force. For periodic boundary conditions one expects the regular part
to be of the order $L^{-4}$ in the vicinity of $T_c$. Much more interesting
is the case of a system with real boundaries. This raises the question of
proper boundary conditions for such systems. They cannot be
boundary conditions of the Dirichlet or Neumann type, because the latter are
incompatible with the long-ranged nature of the interactions. Instead, one
finds that long-ranged interactions generate long-ranged {\em surface
fields} which decay according to a power law away from the surface into the
bulk of the system. The direct (Hamaker) interaction of the surfaces
then generates a $L^{-\sigma}$ contribution to the regular part of the free
energy if the free energy is measured per unit volume (or a $L^{-\sigma+1}$
contribution if the free energy is measured per unit area). This is well
known from studies of, e.g., wetting phenomena \cite{D88}. The contribution
to the regular part of the free energy due to the action of the surface
fields on the ordering degrees of freedom is also
of the order of $L^{-\sigma}$.
The available renormalization group arguments suggest that
the contribution to the singular part should be of the order
$L^{-(d-2+\eta)/2-\sigma}$ \cite{PL83,D86}. This leads to the following
{\it hypothesis for the singular part of the free energy}:
\begin{equation} \label{hypoffss}
f_s(t,L)=L^{-d}X(L/\xi,h_1 L^{\Delta_1/\nu},b
L^{2-\sigma-\eta},h_{s} L^{(d+2-\eta)/2-\sigma}),
\end{equation}
where $h_{s}$ is a nonuniversal metric factor characterizing the
long-ranged behavior of the  surface fields, while $h_1$ is the
corresponding factor at the surface boundaries. Here $\Delta_1$ is
the critical surface gap exponent of the corresponding surface universality
class. Note that, since $d>2$, the term proportional to $h_{s}$
will give contributions that are {\em larger} than that
proportional to $b$, i.e., the contributions of the surface fields
are important and cannot be ignored. In particular, this is
important for the quantitave analysis of wetting experiments with
critical binary liquid mixtures \cite{ML99}. The interpretation of
these experimental data \cite{ML99} is still unresolved. We also
point out, that neither Eq.(\ref{feslr}) nor Eq.(\ref{hypoffss})
applies to confined superfluid $^4$He or $^3$He-$^4$He mixtures
\cite{KD} which have been under investigation in recent
experiments \cite{GC99,GC02}. For superfluid $^4$He the
long-ranged finite-size contributions to the free energy originate
from two distinct sources: (i) a {\em regular} contribution from
the dispersion forces which couple to the fluid {\em density} and
which are unrelated to the superfluid order parameter, (ii) a {\em
singular} contribution from the superfluid order parameter which
is generically short-ranged in nature and does not relate to the
presence of dispersion forces \cite{KD}. The situation is more
complicated in the case of $^3$He-$^4$He mixtures, where the
superfluid transition temperature depends on the $^3$He
concentration, which eventually causes the superfluid transition
to become first order beyond the tricritical $^3$He concentration.
The $^3$He concentration does respond to dispersion forces and, by
virtue of long-ranged surface fields, long-ranged $^3$He
concentration perturbations may emerge in a confined $^3$He-$^4$He
mixture. Through the dependence of the superfluid transition
temperature $T_{\lambda}$ on the $^3$He concentration a
long-ranged variation of the local value of $T_{\lambda}$ will
ensue which in turn imposes a corresponding variation of the
superfluid density in thermal equilibrium. From the experiment in
Ref.\cite{GC02} there is robust evidence that Dirichlet boundary
conditions do not apply for the superfluid order parameter of
$^3$He-$^4$He in the tricritical regime, because contrary to the
case of pure $^4$He \cite{GC99} a thickening of the wetting layer
has been observed caused by a {\em repulsive} tricritical Casimir
force. This observation rules out pure Dirichlet boundary
conditions in this case, because these can only account for {\em
attractive} Casimir forces as observed in pure $^4$He
\cite{KD,GC99}. Furthermore, subdominant long-ranged interactions
may play a significant role in a finite-size scaling analysis of
$^3$He-$^4$He mixtures in the vicinity of the bulk superfluid
transition. However, the model Hamiltonian of the system then has
to accomodate a second 'noncritical' field (the $^3$He
concentration) apart from the superfluid order parameter in order
to include dispersion forces and long-ranged surface fields in the
physically correct way. The construction of a proper model
Hamiltonian, which is a generalization of the standard Ginzburg -
Landau Hamiltonian considered here, is beyond the scope of this
work.

In conclusion, we remark that we still lack a complete theoretical
description of the Casimir effect in systems with subleading long-ranged
interactions. Such a description must contain both the influence of
surface fields and the long-ranged nature of the interaction potential.
Both are expected to generate important contributions to the critical behavior
pertaining to the  universality class of systems with short-ranged interaction
potentials. Very close to $T_c$ the finite-size behavior should turn out to
be that of short-ranged models, but additional finite-size effects are
expected to  become dominant for $|t|L^{1/\nu}\gg 1$. For systems with scalar
order parameter these
expectations hold both above and below the bulk critical temperature. For
$O(n)$ models one expects that Goldstone modes will dominate the finite-size
behavior of the systems below $T_c$.

\acknowledgements
The authors thank Profs. K. Binder, R. Evans, M. E. Fisher, and J. Rudnick
for a critical reading of the manuscript. They are also indebted to one of
the referees for bringing Ref.\cite{C86} to their attention.

D. Dantchev acknowledges the hospitality of Max-Planck-Institute for Metals
Research in Stuttgart as well as the financial support of the Alexander von
Humboldt Foundation.


\begin{thebibliography}{99}
\bibitem{C48} H. B. G. Casimir, Proc. K. Ned. Acad. Wet. {\bf 51},
793 (1948).
\bibitem{C53} H. B. G. Casimir, Physica {\bf 19}, 846 (1953).
\bibitem{MT97} V. M. Mostepanenko and N. N. Trunov, {\em The Casimir Effect and its
Applications} (Oxford University Press, New York, 1997).
\bibitem{M01} K. A. Milton, {\em The Casimir Effect} (World Scientific, Singapore,
2001).
\bibitem{FG78} M. E. Fisher and P. G. de Gennes, C. R. Acad. Sci.
Paris B {\bf 287}, 207 (1978).
\bibitem{K94} M. Krech, {\it The Casimir Effect in Critical Systems}
 (World Scientific, Singapore, 1994).
\bibitem{BDT00} J. G. Brankov, D. M. Danchev, N. S. Tonchev, {\it The
Theory of Critical Phenomena in Finite-Size Systems - Scaling and
Quantum Effects} (World Scientific, Singapore, 2000).
\bibitem{HCM00} B. W. Harris, F. Chen, and U. Mohideen, Phys. Rev. A {\bf 62}, 052109
(2000).
\bibitem{MR98} U. Mohideen and A. Roy, Phys. Rev. Lett.
{\bf 81}, 4549 (1998).
\bibitem{L97} S. K. Lamoreaux, Phys. Rev. Lett. {\bf 78}, 5
(1997).
\bibitem{BSOR02} G. Bressi, G. Carugno, R. Onofrio, and G. Ruoso,
Phys. Rev. Lett. {\bf 88}, 041804 (2002).
\bibitem{L02} A. Lambrecht, Physics World {\bf 15}(9), 29,
(2002).

\bibitem{es}  R. Evans and J. Stecki, Phys. Rev. B {\bf 49}, 8842
(1994).
\bibitem{ParryEvans}  A. O. Parry and R. Evans, Physica A {\bf 181},
250 (1992).
\bibitem{positivneg1}  R. Evans, U. Marini Bettolo Marconi, and P. Tarazona,
J. Chem. Phys. {\bf 84}, 2376 (1986).
\bibitem{positivneg2}  U. Marini Bettolo Marconi, Phys. Rev. A {\bf 38},
6267 (1988).
\bibitem{D88} S. Dietrich, in
{\it Phase Transitions and Critical Phenomena}, edited by C. Domb
and J. L. Lebowitz (Academic, London, 1988), Vol. 12, p. 1.
\bibitem{CD01} X. S. Chen and V. Dohm,  Phys. Rev. E {\bf 66},
016102 (2002).

\bibitem{JJ2002} J. Zinn-Justin, {\em Quantum Field Theory and Critical
Phenomena} (Clarendon, Oxford, 2002).
\bibitem{A84} D. J. Amit, {\em Field Theory, the Renormalization Group, and
Critical Phenomena} (World Scientific, Singapore, 1984).
\bibitem{Fisher65} M. E. Fisher, J. Math. Phys. {\bf 6}, 1643 (1965).
\bibitem{Fisher71} M. E. Fisher, in {\em Critical Phenomena}, Proc. of the
1970 International School of Physics ``Enrico Fermi'', Course LI, edited by
M. S. Green (Academic, New York, 1971), p.1.
\bibitem{PHA91} V. Privman, P. C. Hohenberg, and A. Aharony, in
{\em Phase Transitions and Critical
Phenomena}, edited by C. Domb and J. L. Lebowitz (Academic, London, 1991), Vol. 14, p.1.
\bibitem{FMN72} M. E. Fisher, S.-K. Ma, and B. G. Nickel, Phys.
Rev. Lett. {\bf 29}, 917 (1972).
\bibitem{AF88} M. Aizenman and R. Fern\'{a}ndez, Lett. Math. Phys. {\bf 16}, 39 (1988).
\bibitem{BJG76} E. Br\'{e}zin, J. Zinn - Justin, and J. C. Le Guillou, J. Math. Phys. {\bf 9},
L119 (1976).
\bibitem{FS82} J. Fr\"{o}lich and T. Spencer, Commun. Math. Phys. {\bf 84}, 87 (1982).
\bibitem{C96}J. Cardy, {\em Scaling and Renormalization in Statistical
Physics}  (Cambridge University Press, Cambridge, 1996).
\bibitem{S73} J. Sak, Phys. Rev. B {\bf 8}, 281 (1973).
\bibitem{HN89} J. Honkonen and M. Yu. Nalimov, J. Phys. A {\bf 22}, 751 (1989).
\bibitem{H90} J. Honkonen, J. Phys. A {\bf 23}, 825 (1990).
\bibitem{J98} H. K. Janssen, Phys. Rev. E {\bf 58}, R2673 (1998).
\bibitem{LB2002} E. Luijten and H. W. J. Bl\"{o}te, Phys. Rev. Lett. {\bf 89},
025703 (2002).
\bibitem{CEHH94} R. J. F. Leote de Carvalho, R. Evans, D. C. Hoyle and J. R. Henderson,
J. Phys. : Cond. Matt. {\bf 6}, 9275 (1994).
\bibitem{KR84} R. F. Kayser and H. J. Ravech\'e,  Phys. Rev. A
  {\bf 29}, 1013 (1984).
\bibitem{FD95} G. Fl\"oter and S. Dietrich,  Z. Phys. B {\bf 97},
  213 (1995).
\bibitem{D01} D. M. Dantchev, Eur. Phys. J. B. {\bf 23}, 211 (2001).
\bibitem{CW73} B. M. McCoy and T. T. Wu, {\em The Two-dimensional Ising Model}
(Harvard University Press, Cambridge, 1973).
\bibitem{FB67} M. E. Fisher and R. J. Burford,  Phys. Rev.  {\bf
    156}, 583 (1967).
\bibitem{CD2000} X. S. Chen and V. Dohm,  Eur. Phys. J. {\bf B 15},
283 (2000).
\bibitem{Fisher72} M. E.  Fisher, in {\it Critical Phenomena}, Proc. 51st
Enrico Fermi Summer School, Varenna, edited by M. S. Green
(Academic, New York, 1972), p. 1.
\bibitem{fisherbarber72} M. E. Fisher and M. Barber, Phys. Rev. Lett. {\bf 28},
1516 (1972)
\bibitem{barber83} M. Barber, in {\it Phase Transitions and Critical
Phenomena}, edited by C. Domb and J. L. Lebowitz (Academic
Press, New York, 1983), Vol. 8, p. 145.
\bibitem{PF84} V. Privman and M. E. Fisher, Phys. Rev. B {\bf 30}, 322 (1984).
\bibitem{privman90} V. Privman, in {\it Finite Size Scaling and Numerical
Simulation of Statistical Systems}, edited by V. Privman (World
Scientific, Singapore, 1990), p. 1.

\bibitem{DR01} D. Dantchev and J. Rudnick, Eur. Phys. J.
B {\bf 21}, 251 (2001).

\bibitem{rem1} Here  $\xi\equiv\xi(L,t)$ is the finite-size correlation length, see, e.g.,
\cite{BDT00}; $\xi(L\rightarrow\infty,t)\rightarrow\xi(\infty,t)$, with
$\xi(\infty,t)\simeq \xi_0^+t^{-\nu}$, where $\nu=1/(d-2)$.
\bibitem{KD} M. Krech and S. Dietrich, Phys. Rev. Lett. {\bf 66}, 345 (1991);
Phys. Rev. A {\bf 46}, 1886 (1992); {\bf 46}, 1922 (1992).

\bibitem{K99} M. Krech, J. Phys.: Condens. Matter {\bf 11}, R391 (1999).
\bibitem{ML99} A. Mukhopadhyay and B. M. Law, Phys. Rev. Lett. {\bf 83}, 772
(1999); Phys. Rev. E {\bf 62}, 5201 (2000).
\bibitem{GC99} R. Garcia and M. H. W. Chan, Phys. Rev. Lett. {\bf 83}, 1187
(1999).
\bibitem{GC02} R. Garcia and M. H. W. Chan, Phys. Rev. Lett. {\bf 88}, 086101
(2002).
\bibitem{CD99} X. S. Chen and V. Dohm, Eur. Phys. J. B
 {\bf 7}, 183 (1999).
\bibitem{CD} X. S. Chen and V. Dohm, Eur. Phys. J. B
 {\bf 10}, 687 (1999).
\bibitem{C86} J. L. Cardy, in {\it Fields, Strings and Critical Phenomena}, edited by E. Br\'{e}zin
and J. Zinn-Justin (Elsevier, 1989), p. 169.
\bibitem{CDcm} X. S. Chen and V. Dohm, cond-mat/0108202.
\bibitem{P92} G. Parisi, {\it Statistical Field Theory} (Addison-Wesley, Reading, MA, 1992).
\bibitem{rem2} See ``Note added in proof'', p. 700,  of Ref. \cite{CD}.
\bibitem{HD02} H. Chamati and D. M. Dantchev, Eur. Phys. J. B. {\bf 26},
 89 (2002).
\bibitem{FP84} V. Privman and M. E. Fisher, Phys. Rev. B {\bf  30}, 322 (1984).

\bibitem{PL83} L. Peliti and S. Leibler, J. Phys. C.: Solid State Phys. {\bf 16},
2635 (1983).
\bibitem{D86} H. W. Diehl, in {\it Phase Transitions and Critical
Phenomena}, edited by
C. Domb and J. L. Lebowitz (Academic, London, 1986), Vol. 10, p. 76.
\end{thebibliography}
\end{document}